\documentclass[12pt]{article}

\usepackage{graphics,amssymb,epsfig,float}
\usepackage[usenames,dvips]{color}
\usepackage{graphicx}
\usepackage{epsfig}
\usepackage{rotating}
\usepackage{dcolumn}
\usepackage{bm}
\usepackage{cite}
\usepackage{amsmath}
\usepackage{setspace}

\def\lsim{\;\raise0.3ex\hbox{$<$\kern-0.75em\raise-1.1ex\hbox{$\sim$}}\;}
\def\gsim{\;\raise0.3ex\hbox{$>$\kern-0.75em\raise-1.1ex\hbox{$\sim$}}\;}
\def\beq{\begin{equation}}   \def\eeq{\end{equation}}
\def\ba{\begin{array}}       \def\ea{\end{array}}
\def\bea{\begin{eqnarray}}   \def\eea{\end{eqnarray}}
\def\nn{\nonumber}

\begin{document}

\begin{titlepage}
\begin{flushright}
LPT Orsay 13-50
\end{flushright}

\begin{center}
\vspace{1cm}
{\Large\bf Higgs pair production in the NMSSM at the LHC}
\vspace{2cm}

{\bf{Ulrich Ellwanger}}
\vspace{1cm}\\
Laboratoire de Physique Th\'eorique, UMR 8627,\\ CNRS and
Universit\'e de Paris--Sud,
F-91405 Orsay, France

\end{center}

\vspace{1cm}

\begin{abstract}
In the NMSSM it is well possible to find an additional Higgs boson with
a mass below 125~GeV which remains invisible in standard Higgs boson
search channels. We study the Higgs pair production cross sections times
branching fractions in this scenario, focusing on gluon fusion and the
$b\bar{b}+\tau^+\tau^-$ and $b\bar{b}+\gamma\gamma$ final states.
Summing over the SM-like and the lighter Higgs states, the production
cross sections times branching fractions are never below the ones for SM
Higgs pair production. Sizable enhancements of the signal rates are
also possible, notably if a lighter Higgs state is produced. However,
the rates involving at least one lighter Higgs boson are not
always sufficiently large to guarantee its discovery.
\end{abstract}

\end{titlepage}

\newpage
\section{Introduction}

The couplings of the 125~GeV Higgs boson to electroweak gauge bosons,
third generation fermions and the loop induced couplings to gluons and
photons have been measured by the ATLAS and CMS collaborations with an
astonishing precision already after the 7-8~TeV runs
\cite{ATLAS-CONF-2013-034,CMS-PAS-HIG-13-005}. Clearly, these
measurements impose constraints on models with extended Higgs sectors.

However, extended Higgs sectors can manifest themselves mainly through
deviations of Higgs self couplings from the values within the Standard
Model (SM), which can now be determined once the Higgs mass is known.
For this reason, measurements of Higgs self couplings (actually only the
trilinear self coupling seems accessible in the foreseeable future) at
the LHC at 14~TeV at high integrated luminosity and/or at future
colliders are of utmost importance.

The trilinear Higgs self coupling contributes to Higgs pair production
which allows, in principle, its measurement. Many
corresponding studies have been performed in the SM and its
supersymmetric (Susy) extensions, notably at hadron colliders and
considering gluon fusion
\cite
{Eboli:1987dy,Dicus:1987ic,Glover:1987nx,Plehn:1996wb,Dawson:1998py,
Djouadi:1999rca, Belyaev:1999mx,Belyaev:1999vz,Cynolter:1999bn,
Kim:2000rv,BarrientosBendezu:2001di, Baur:2002rb,Baur:2002qd,
Baur:2003gpa, Baur:2003gp, Binoth:2006ym,Dolan:2012rv,
Christensen:2012si, Papaefstathiou:2012qe, Dolan:2012ac, Baglio:2012np,
Shao:2013bz,Goertz:2013kp, Cao:2013si,deFlorian:2013uza,
Gouzevitch:2013qca,Grigo:2013rya, Nhung:2013lpa}, the
production mode with the largest cross section.

Among the Susy extensions of the SM, the Next-to-Minimal Supersymmetric
Standard Model (NMSSM) \cite{Ellwanger:2009dp} has received considerable
attention \cite
{Hall:2011aa,Ellwanger:2011aa,Arvanitaki:2011ck,King:2012is,Kang:2012sy,
Cao:2012fz,Ellwanger:2012ke,Jeong:2012ma,Randall:2012dm,Benbrik:2012rm,
Kyae:2012rv,Cao:2012yn,Agashe:2012zq,Belanger:2012tt,Heng:2012at,
Choi:2012he,King:2012tr,Gherghetta:2012gb,Cheng:2013fma,
Barbieri:2013hxa,Badziak:2013bda,Cheng:2013hna,Hardy:2013ywa},
since a Higgs mass of 125~GeV is more natural within its parameter
space than in the MSSM. Due to
the additional gauge singlet superfield $S$ compared to the MSSM, the
NMSSM contains three neutral CP-even Higgs states $H_i, i=1,2,3$
(ordered in mass). 

It is quite natural in the parameter space of the NMSSM that the mostly
SM-like state $H_{\text{SM}}$ near 125~GeV is actually the second
lightest state $H_2$, wheras the lightest state $H_1$ is mostly
singlet-like; then mixing effects contribute to the increase of the mass
of $H_{\text{SM}}$. A lighter mostly singlet-like state $H_1$ can well
be compatible with the constraints from LEP \cite{Schael:2006cr}
(simultaneously with a SM-like state near 125~GeV
\cite{Belanger:2012tt}), and might even explain the mild ($\sim
2\sigma$) excess in $e^+ e^- \to Zb\bar{b}$ near $M_{b\bar{b}}\sim
100$~GeV \cite{Schael:2006cr}.

The pair production of two states $H_{\text{SM}}$ in the NMSSM has
been considered recently in \cite{Cao:2013si,Nhung:2013lpa}. In
\cite{Cao:2013si} it was found that, in the presence of light stops
present in the loop-induced amplitudes, the production rate can be
considerably larger than in the SM. In \cite{Nhung:2013lpa} it was
emphasized that loop corrections to the trilinear Higgs couplings and
hence to the pair production of two states $H_{\text{SM}}$ can
potentially be large. (The range of the trilinear Higgs couplings within
some regions of the parameter space of the NMSSM have been studied in
\cite{Gupta:2013zza}.)

In the present paper we concentrate on the case where the mostly
SM-like state $H_{\text{SM}}$ near 125~GeV corresponds to $H_2$, and
study the prospects of $H_2+H_2$, $H_2+H_1$ and $H_1+H_1$ pair
production. Notably in the case of a small non-singlet component of
$H_1$, $H_1$ will hardly couple to electroweak gauge bosons and
fermions, and be practically invisible in standard Higgs search
channels. However, the NMSSM-specific $H_1 H_2 H_2$ and/or $H_1 H_1 H_2$
couplings can still be large. These would allow to detect $H_1$ in $H_2
\to H_1+H_1$ decays if $M_{H_1} \lsim M_{H_2}/2$. For $M_{H_1} \gsim
M_{H_2}/2$, the case considered here, Higgs pair production might be the
only way to observe the $H_1$ state.

In principle $H_1$ could also be produced in $H_3$ decays
\cite{Kang:2013rj,Barbieri:2013hxa}. However, this strategy would also
fail if $H_3$ is too heavy. Subsequently we make the pessimistic
assumption that this is the case, and that $H_3$ does not contribute to
the Higgs pair production cross section. Likewise we assume that stops
are too heavy to affect the Higgs pair production cross section;
otherwise stops are likely to be discovered before the observation of
Higgs pair production processes. Also light pseudoscalars are assumed to
be absent. If the assumptions of a heavy $H_3$, heavy pseudoscalars and
heavy stops turn out to be wrong, we would be pleased to redo the
calculations with correspondingly known masses.

At present the most promising search strategy for Higgs pair production
at the LHC at 14~TeV seems the application of subjet-based analysis
techniques to boosted kinematical regimes of the dihiggs system
\cite{Dolan:2012rv}. According to \cite{Dolan:2012rv}, the
$b\bar{b}+\tau^+\tau^-$ final state seems accessible by this method. In
the case of the NMSSM, the branching fractions into $b\bar{b}$ and
$\tau^+\tau^-$ can differ from the SM. Notably the lighter state $H_1$
will hardly decay into electroweak gauge bosons $W^\pm$ and $Z$, which
implies a somewhat larger branching fraction into $b\bar{b}$ (and
$\tau^+\tau^-$) than the $\sim 60\%$ of a SM-like Higgs boson. The
corresponding branching fractions of $H_2$ can well be reduced. For
these reasons we will study the Higgs pair production cross sections
multiplied by the corresponding branching fractions normalized to the
SM.

The $b\bar{b}+W^+W^-$ final state in boosted kinematical regimes has
been analysed as well using jet substructure techniques
\cite{Papaefstathiou:2012qe}. However, due to the reduced branching
fractions of $H_1$ into $W^+W^-$ we will not consider this channel here.
More recently, analyses of the $b\bar{b}+b\bar{b}$ final state
\cite{Gouzevitch:2013qca} have been proposed as promising in the case of
Higgs pair production via heavy resonances.

Another possible final state is $b\bar{b}+\gamma\gamma$
\cite{Baur:2003gp}. In spite of the limited statistics, Higgs pair
production may be observable in this channel if fake $b$-jets and
photons are sufficiently under control. Hence we extend our analysis to
this final state, again multiplying the Higgs pair production cross
sections by the corresponding branching fractions normalized to the~SM.

In the case of the state $H_2$ near 125~GeV, the measured couplings
\cite{ATLAS-CONF-2013-034,CMS-PAS-HIG-13-005} imply bounds on possible
deviations from the SM. Here we will apply bounds (at the 95\%
confidence level) on all couplings from a recent combination of the
ATLAS, CMS and Tevatron results \cite{Belanger:2013xza}. These constrain
not only the $H_2$ branching fractions and couplings, but all NMSSM
specific parameters.

In the next Section we describe our calculations, in Section~4 our
results. Conclusions are devoted to Section~3.

\section{Calculations of Higgs pair production cross
sections in the NMSSM}

In any model for the Higgs sector, the effective potential describes the
Higgs vacuum expectation values (vevs), the (running) Higgs masses and
Higgs self couplings. Radiative corrections to the effective potential
will affect all these quantities simultaneously. Given the knowledge of
the values of the $SU(2)\times U(1)$ symmetry breaking vev and the
SM-like Higgs mass, it is crucial to compute the Higgs self couplings
at the same level of precision as the SM-like Higgs mass.

For a given set of input parameters (see below), the radiative
corrections to the Higgs masses and couplings in the NMSSM can be
sizable \cite{Ellwanger:2009dp,Nhung:2013lpa}. For not too light Higgs
states circulating in the loops (as considered here), the dominant
radiative corrections to the effective potential originate from stop/top
quark loops. Since we scan over the input parameters of the general
$\mathbb{Z}_3$-invariant NMSSM (and, to some extent, radiative
corrections to couplings can be absorbed by modifications of the input
parameters such that the SM-like Higgs mass remains unchanged) we
confine ourselves to the dominant radiative corrections from stop/top
quark loops to masses and Higgs self couplings in the leading
logarithmic approximation. To this end we employ a correspondingly
modified version of the code NMSSMTools
\cite{Ellwanger:2004xm,Ellwanger:2005dv,nmweb}.

The field content of the Higgs sector of the NMSSM are two SU(2)
doublets $H_u$, $H_d$, and a gauge singlet $S$. The parameters in the
Higgs sector of the general $\mathbb{Z}_3$-invariant NMSSM include two
dimensionless Yukawa couplings $\lambda$ (proportional to a term $S H_u
H_d$ in the superpotential) and $\kappa$ (a trilinear singlet self
coupling in the superpotential), corresponding soft Susy breaking terms
$A_\lambda$, $A_\kappa$, and three soft Susy breaking mass terms for
$H_u$, $H_d$, and $S$. The three soft Susy breaking mass terms can be
traded for the known value of $M_Z$ and the two variables $\tan\beta
\equiv \left<H_u\right>/\left<H_d\right>$, $\mu_\text{eff} \equiv
\lambda\left<S\right>$ (for more details see \cite{Ellwanger:2009dp}).
Hence we are left with the free variables
\beq
\lambda,\ \kappa,\ A_\lambda,\ A_\kappa,\ \tan\beta\ \text{and}\
\mu_\text{eff}\; .
\eeq

A mass of about 125~GeV of a mostly SM-like Higgs boson $H_2$ is
obtained naturally for larger values of $\lambda$ and relatively low
values of $\tan\beta$. The results below are obtained from a scan over
these parameters in the ranges
\bea
&\lambda=0.5-0.7,\quad &\kappa=0.07-0.5,\nn\\
&A_\lambda=300-1000~\text{GeV},\quad &A_\kappa= -500-0~\text{GeV},\nn\\
&\tan\beta=1.5-6,\quad &\mu_{\text{eff}}=120-300~\text{GeV}\; .
\eea

Since we confine ourselves to the leading stop/top quark induced
radiative corrections to the Higgs potential, we only have to specify
the stop masses and mixing parameter. We choose 1~TeV for the stop
masses (allowing to safely neglect stop contributions to Higgs
production in gluon fusion) and $A_\text{top}=0$ for the mixing
parameter. (Large values for $A_\text{top}$ are not necessary in order
to obtain a mostly SM-like Higgs boson at 125~GeV through radiative
corrections.)

The Higgs spectrum in the above range of parameters contains no light
CP-odd states. Together with the relatively low values of $\tan\beta$,
the constraints from ${\cal{B}}$-physics implemented in NMSSMTools are
always satisfied. More relevant are the constraints we impose on the
CP-even Higgs sector.

First, we require a CP-even Higgs boson $H_2$ of a mass of $125.7 \pm
2$~GeV. Next we require that the signal rates of this state in the
production modes via gluon fusion, vector boson fusion and in
association with electroweak gauge bosons, and in the decay channels
into two photons, electroweak gauge bosons, pairs of $b$-quarks and
$\tau$-leptons are within the 95\% confidence limits obtained in
\cite{Belanger:2013xza} from a combination of the measurements at ATLAS,
CMS and the Tevatron. (In the NMSSM at low $\tan\beta$ one can safely
assume that the Higgs couplings to $b$-quarks and $\tau$-leptons are
rescaled by the same amount with respect to the SM, so that the
corresponding measurements can be combined.)

These bounds imply that, within the above parameter space of the NMSSM,
the couplings of the Higgs boson near 125~GeV to electroweak gauge
bosons and fermions cannot deviate dramatically from the SM values, just
the loop-induced coupling to photons can be significantly larger
(depending on the production mode, see \cite{Belanger:2013xza}).

On the mass of the lightest CP-even Higgs boson $H_1$ we impose $M_{H_1}
> 65$~GeV.  Otherwise $H_2 \to H_1+H_1$ decays (with $H_2$ nearly on
shell) would be possible with much larger signal rates than the Higgs
pair production considered here, allowing to detect $H_1$ in single
$H_2$ production channels. Subsequently we assume that such on-shell
Higgs-to-Higgs decays will not be observed.

The code NMSSMTools provides us with all required parameters (including
the trilinear Higgs self couplings) for the calculation of the Higgs
pair production cross sections in gluon fusion of all possible
combinations of $H_1$ and $H_2$ states. For the calculation of the Higgs
pair production cross sections we use the public
code {\sf HPAIR} \cite{hpair} which includes QCD corrections and a low
energy theorem for the top quark loop. We modified this code according
to our needs and verified that, in cases where $H_1$ or $H_2$ are
SM-like and the other state is omitted, a SM-like pair production cross
section is obtained. (For a 125~GeV SM Higgs boson, {\sf HPAIR} gives a
production cross section of about $32.6$~fb in gluon fusion at 14~TeV
c.~m. energy.)

\section{Results for Higgs pair production cross sections in the NMSSM}

First we show our results for the trilinear Higgs couplings obtained
from NMSSMTools from the scan described in Section~2. These are defined
in the convention where the the trilinear terms in the effective
potential are written as
\beq
V_{\text{eff}}= \frac{1}{3!} \sum_{i,j,k} g(H_i,H_j,H_k) H_i H_j H_k
+ ...\; ,
\eeq
in which the trilinear Higgs coupling of the SM has a value of about
190~GeV. Our subsequent results will be shown as function of $M_{H_1}$
which we study in the range $M_{H_1}\sim 65-125$~GeV (always below the
SM-like $M_{H_2}$).
In Fig.~1 we show scatter plots of $g(H_2,H_2,H_2)$, $g(H_1,H_2,H_2)$,
$g(H_1,H_1,H_2)$ and $g(H_1,H_1,H_1)$.

\begin{figure}[ht!]\centering
\includegraphics[width=6.1cm,angle=-90]{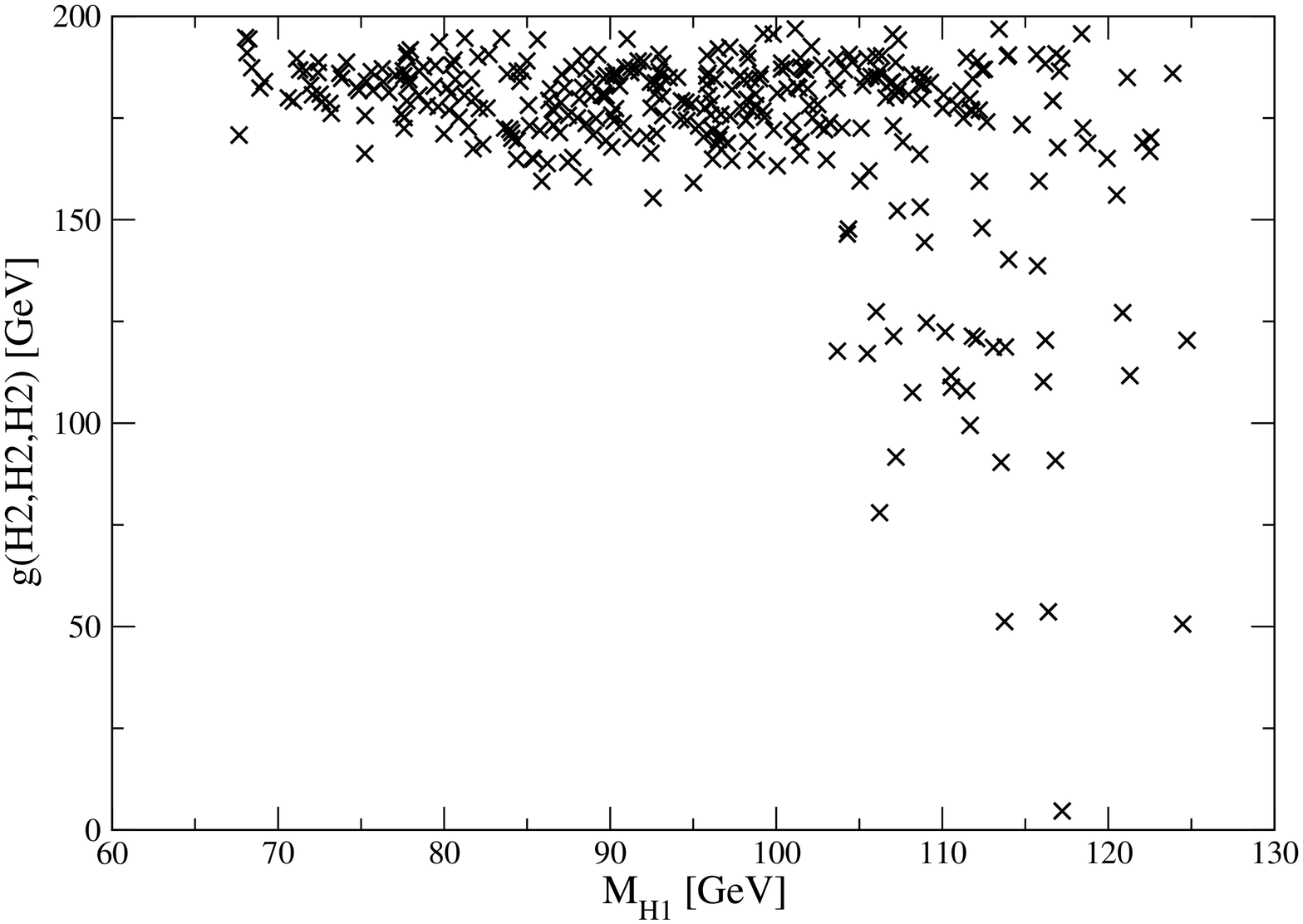}\quad
\includegraphics[width=6.1cm,angle=-90]{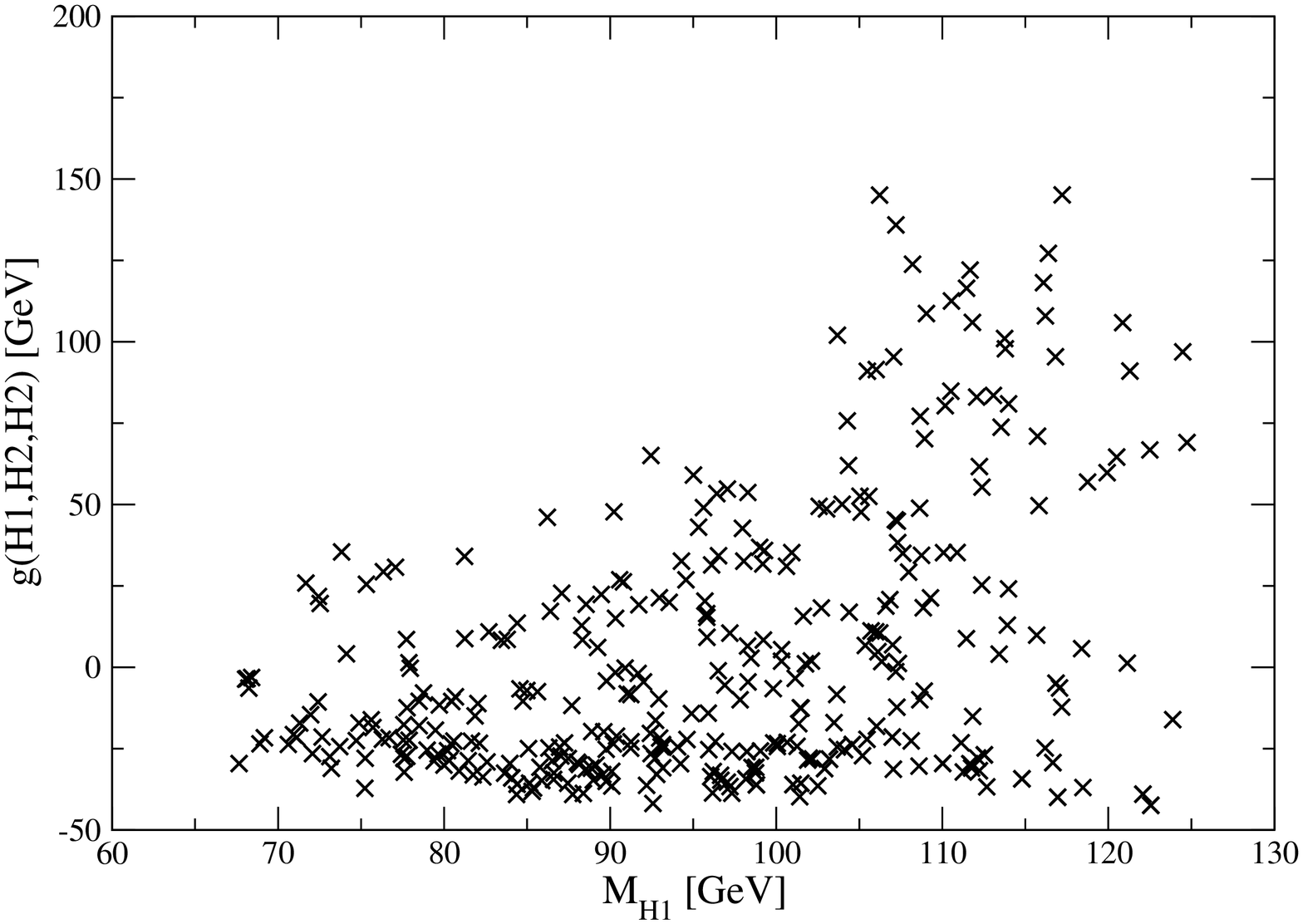}\\
\includegraphics[width=6.1cm,angle=-90]{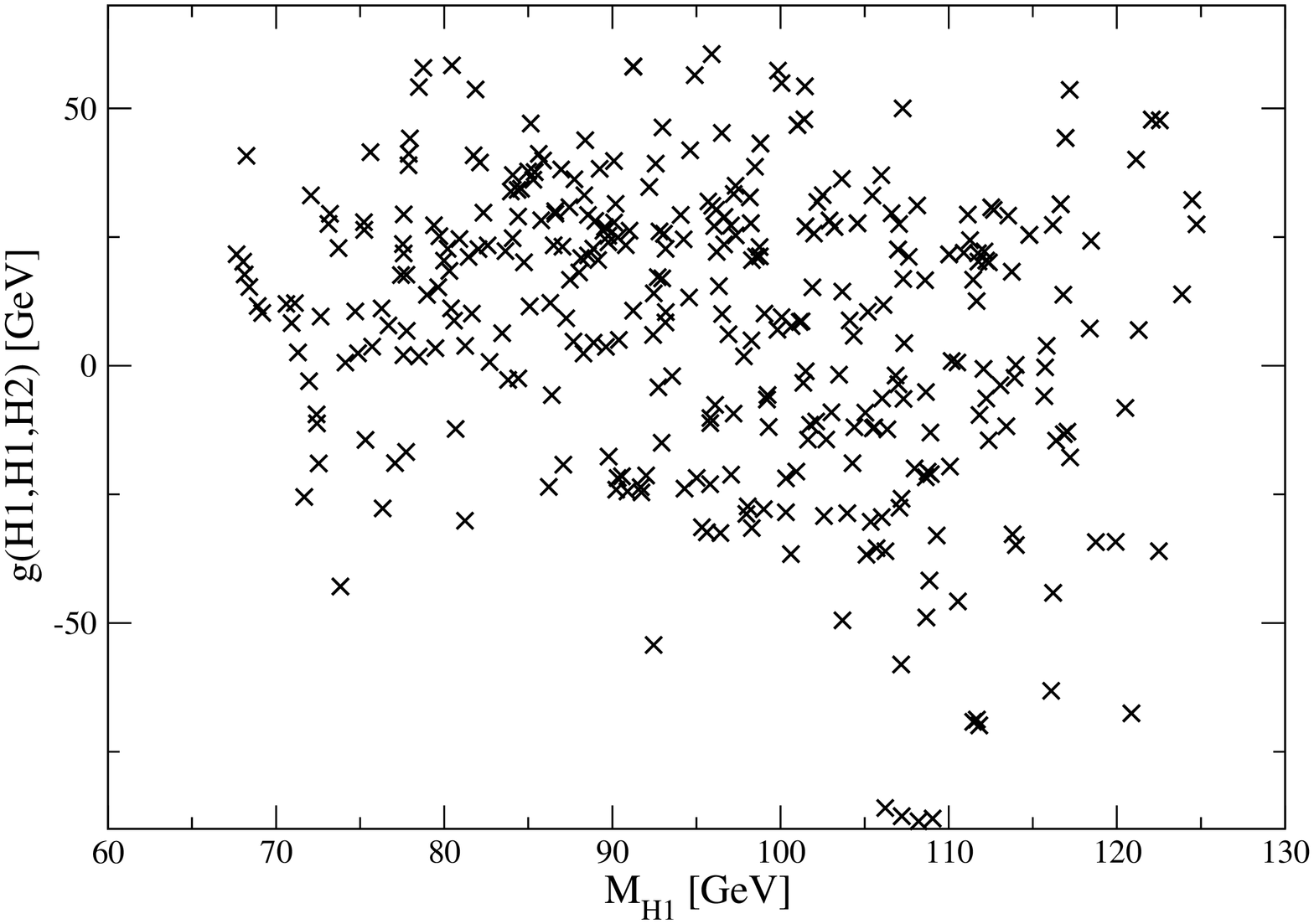}\quad
\includegraphics[width=6.1cm,angle=-90]{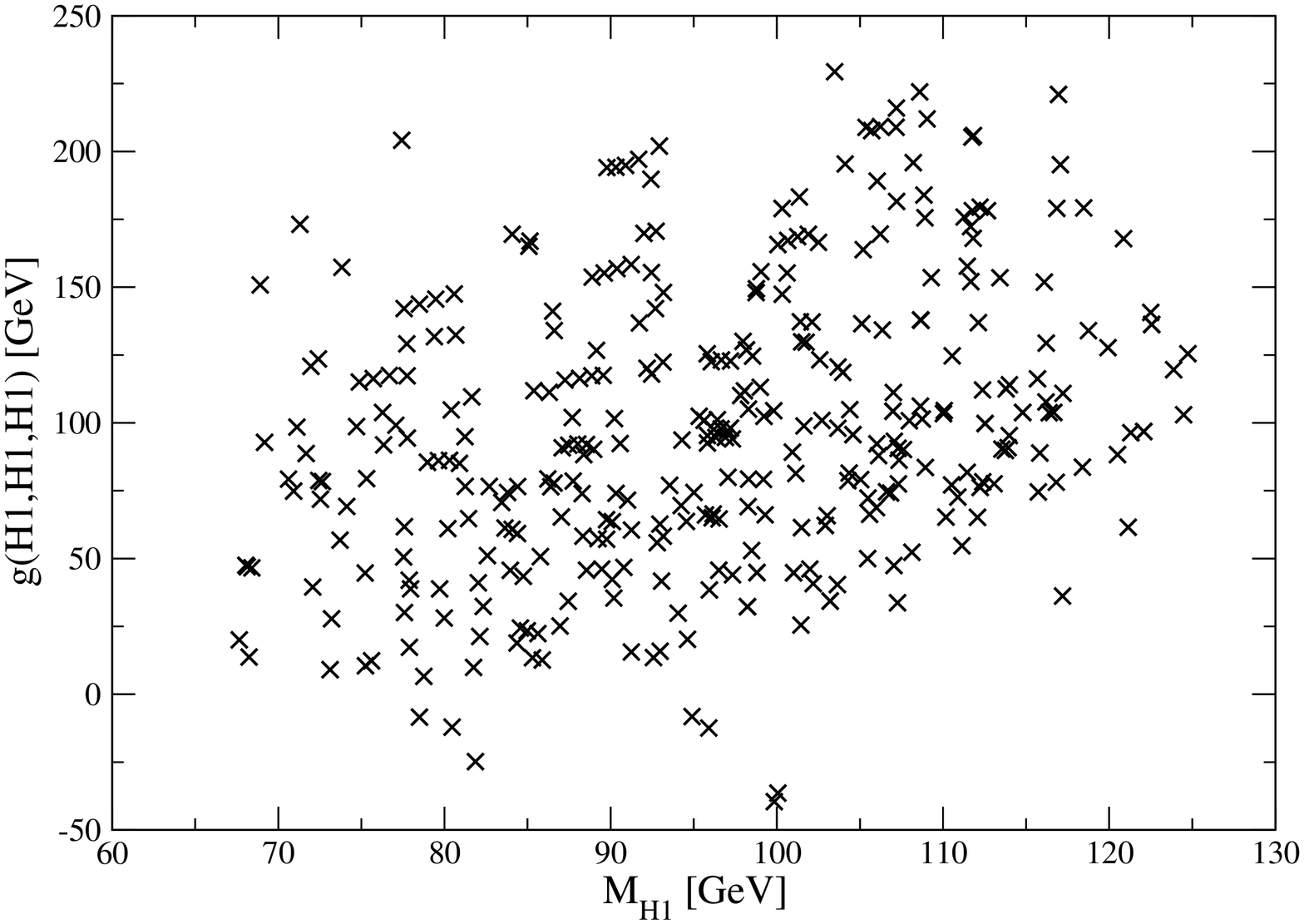}
\caption{Scatter plots of the trilinear Higgs couplings
$g(H_2,H_2,H_2)$ (top left), $g(H_1,H_2,H_2)$ (top right),
$g(H_1,H_1,H_2)$ (bottom left) and $g(H_1,H_1,H_1)$ (bottom right) as
function of $M_{H_1}$.}
\label{fig:1}
\end{figure}

We see that the trilinear coupling of the SM-like state $H_2$ (top left)
is typically close to its SM value, except for a possible reduction if
$M_{H_1} \gsim 100$~GeV. The couplings $g(H_1,H_2,H_2)$ and
$g(H_1,H_1,H_2)$ are smaller, just $g(H_1,H_2,H_2)$ can assume similar
values as $g(H_2,H_2,H_2)$ for $M_{H_1} \gsim 100$~GeV. The coupling
$g(H_1,H_1,H_1)$ can vary from small up to SM-like values, with no
specific dependence on $M_{H_1}$. Values significantly larger than in
the SM are not observed.

Next we turn to the Higgs pair production cross sections in gluon fusion
at the LHC at 14~TeV. As stated in the Introduction, we make the
conservative assumptions that neither light stops nor the heavy state
$H_3$ contribute to the cross sections. Since observations of Higgs pair
production will certainly require large integrated luminosities of at
least 300-600~fb$^{-1}$, these particles should otherwise have been
observed by then in which case the present calculations can be
correspondingly corrected. Besides the triple Higgs couplings shown
before, also the Higgs top couplings are relevant here.

We normalize all Higgs pair production cross sections to the SM value of
about 32.6~fb from {\sf HPAIR}. First, this value will certainly be
used as ``benchmark'' by the experimental collaborations in order to
test the trilinear Higgs coupling of the SM. Second, one can expect that
this ratio should remain approximately invariant under higher order QCD
corrections and the replacement of the heavy top quark limit in {\sf
HPAIR} by the full matrix element, although the latter may be relevant
for specific simulations \cite{Dolan:2012rv}.
In Fig.~2 we show the normalized Higgs pair production cross sections
for the final states $H_2+H_2$, $H_2+H_1$, $H_1+H_1$ and the sum over
all final states.

\begin{figure}[ht!]\centering
\includegraphics[width=6.1cm,angle=-90]{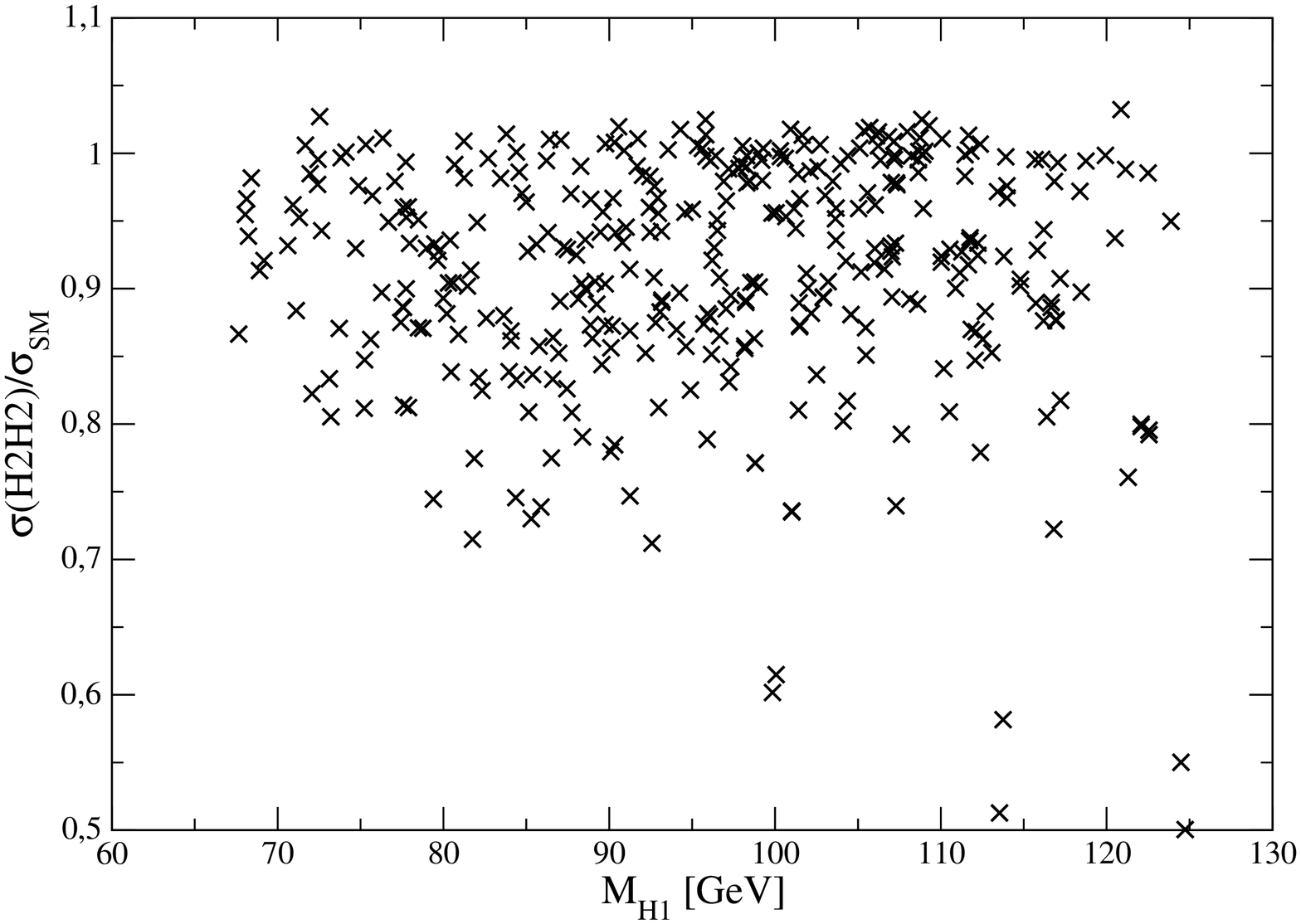}\quad
\includegraphics[width=6.1cm,angle=-90]{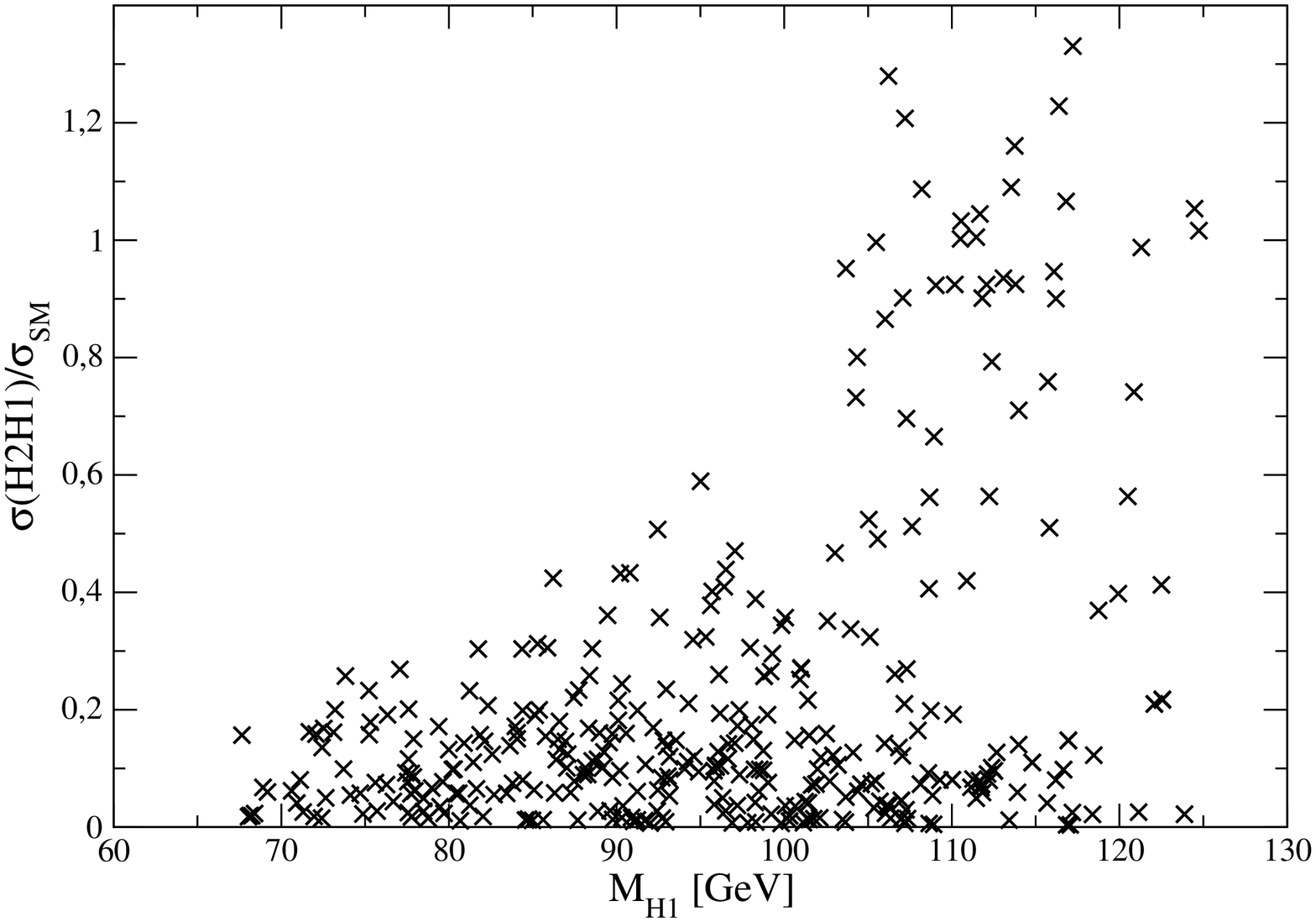}\\
\includegraphics[width=6.1cm,angle=-90]{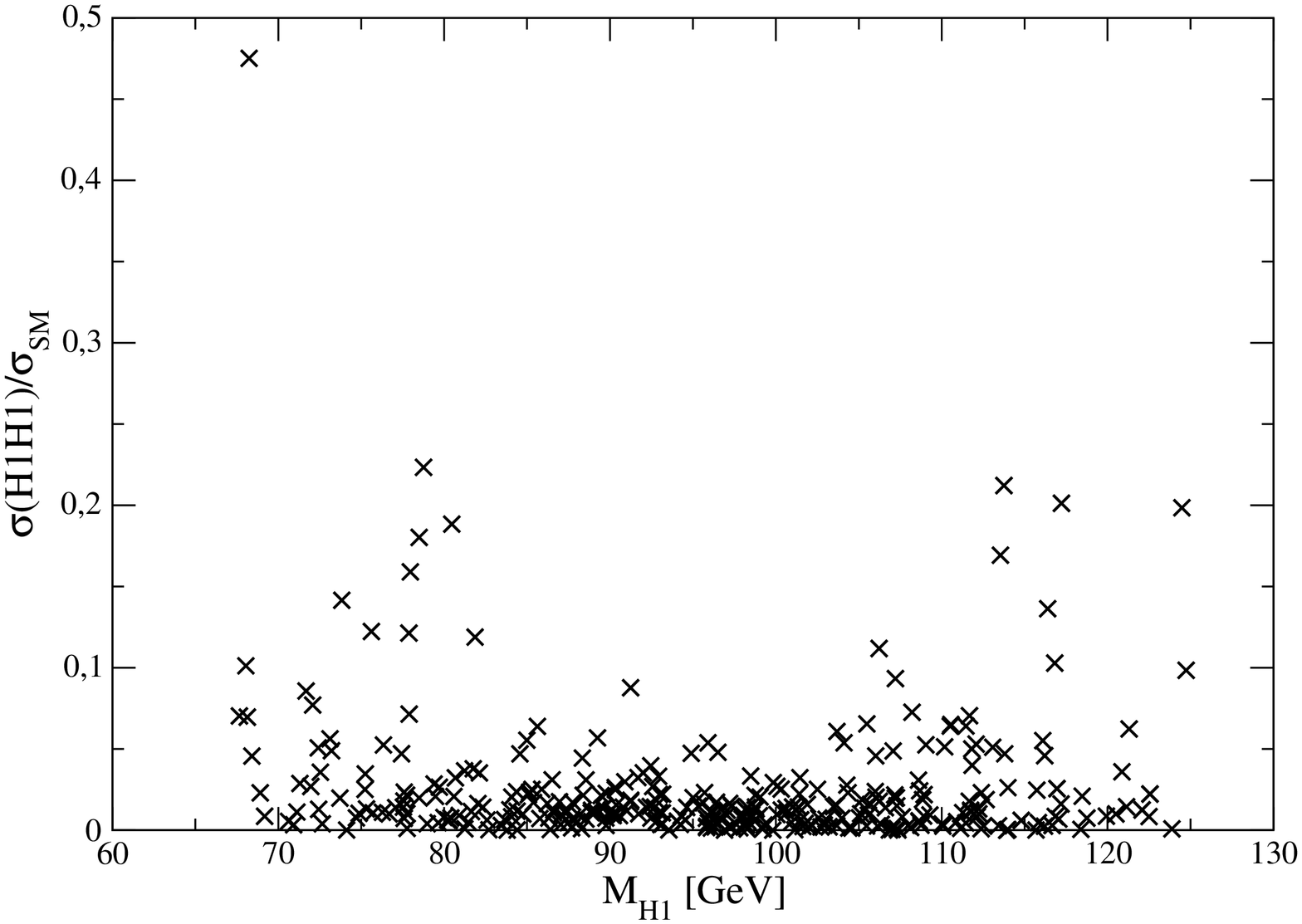}\quad
\includegraphics[width=6.1cm,angle=-90]{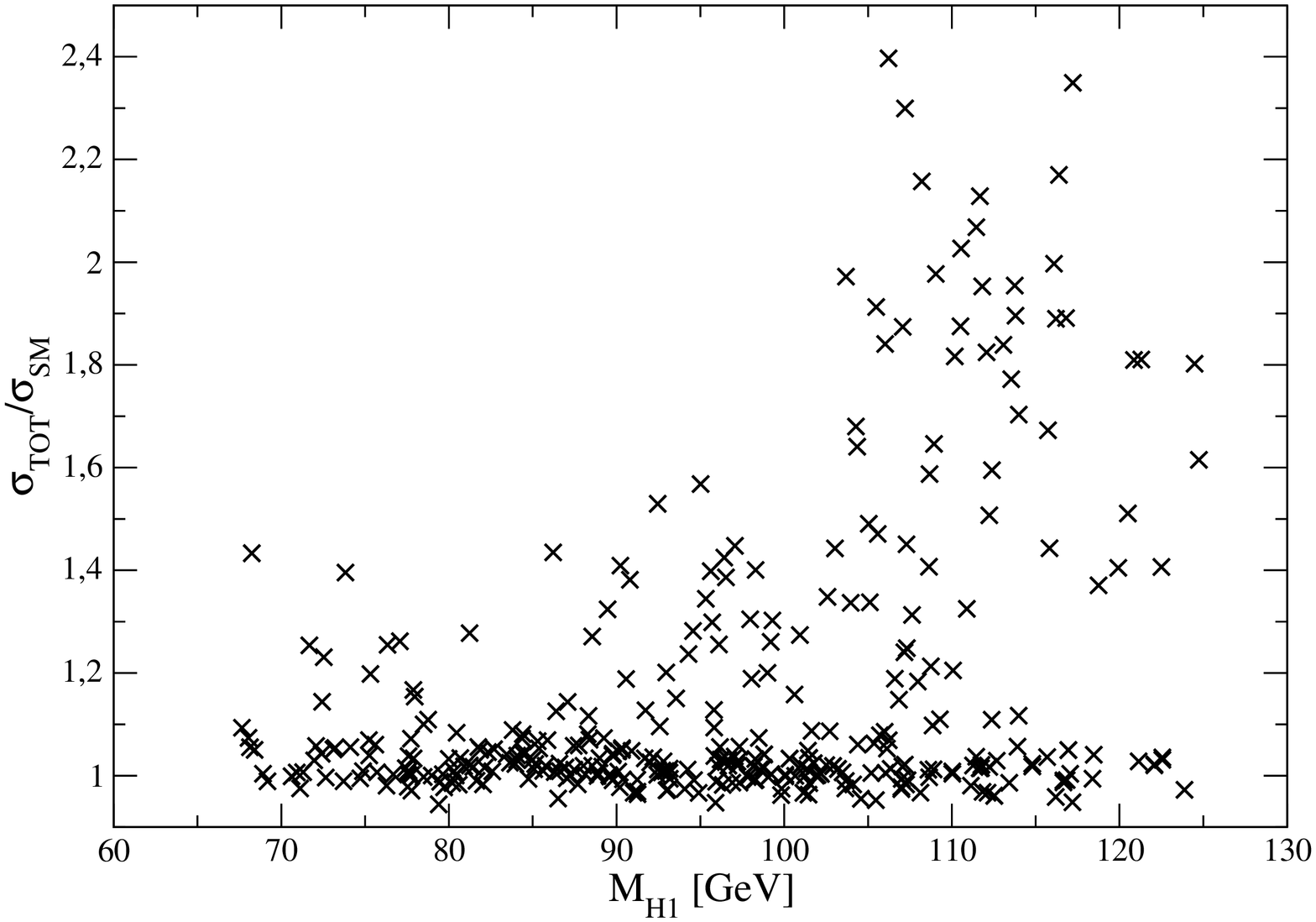}
\caption{The Higgs pair production cross sections relative to the SM
for the final states $H_2+H_2$ (top left), $H_2+H_1$ (top right),
$H_1+H_1$ (bottom left) and the sum over
all final states (bottom right).}
\label{fig:2}
\end{figure}

In Fig.~2 we see no significant enhancement of the cross section for the
final state $H_2+H_2$ (in agreement with the results obtained in
\cite{Cao:2013si} for heavy stops), a possible enhancement for the final
state $H_2+H_1$ for $M_{H_1} \gsim 100$~GeV, and always below-the-SM
values for the final state $H_1+H_1$. Unfortunately there exist regions
in the parameter space where the cross sections for both final states
$H_2+H_1$ and $H_1+H_1$ are small. Since $H_1$ has typically reduced
couplings to top quarks, its production has to rely essentially on a
virtual mostly SM-like $H_2$ in the s-channel. However, the then
relevant trilinear couplings $g(H_1,H_2,H_2)$ and $g(H_1,H_1,H_2)$ can
simultaneously be small leading to a poor production rate for both
$H_2+H_1$ and $H_1+H_1$ final states, which are relevant for the $H_1$
discovery. The good news is, on the other hand, that the sum over all
final states is never below the SM value, but up to $\sim 2.5$ times
larger for $M_{H_1} \gsim 100$~GeV.

However, Higgs pair production will only be measurable in specific
channels. First we consider the $b\bar{b}+\tau^+\tau^-$ final state
which has been studied in \cite{Dolan:2012rv}. Since the branching
fractions of both $H_1$ and $H_2$ can differ significantly from  a
SM-like Higgs boson (limited, in the case of $H_2$, by the
combinations of the LHC and Tevatron measurements) one should weight the
production cross sections by the corresponding branching fractions
relative to the SM. In the NMSSM for low $\tan\beta$ one can assume that
the relative variations of the branching fractions into $b\bar{b}$ and
$\tau^+\tau^-$ are practically the same, since both couplings originate
from the $H_d$ component of $H_1$ and $H_2$. Hence, in the case of
$H_2+H_1$, these cross sections weighted by the branching fractions will
be the same for $H_2$ or $H_1$ decaying into~$\tau^+\tau^-$.

In Fig.~3 we show the normalized Higgs pair production cross sections
for the final states $H_2+H_2 \to b\bar{b}+\tau^+\tau^-$, $H_2+H_1 \to
b\bar{b}+\tau^+\tau^-$, $H_1+H_1 \to b\bar{b}+\tau^+\tau^-$ and the sum
over all $H_i$ into $b\bar{b}+\tau^+\tau^-$.

\begin{figure}[t!]\centering
\includegraphics[width=6.1cm,angle=-90]{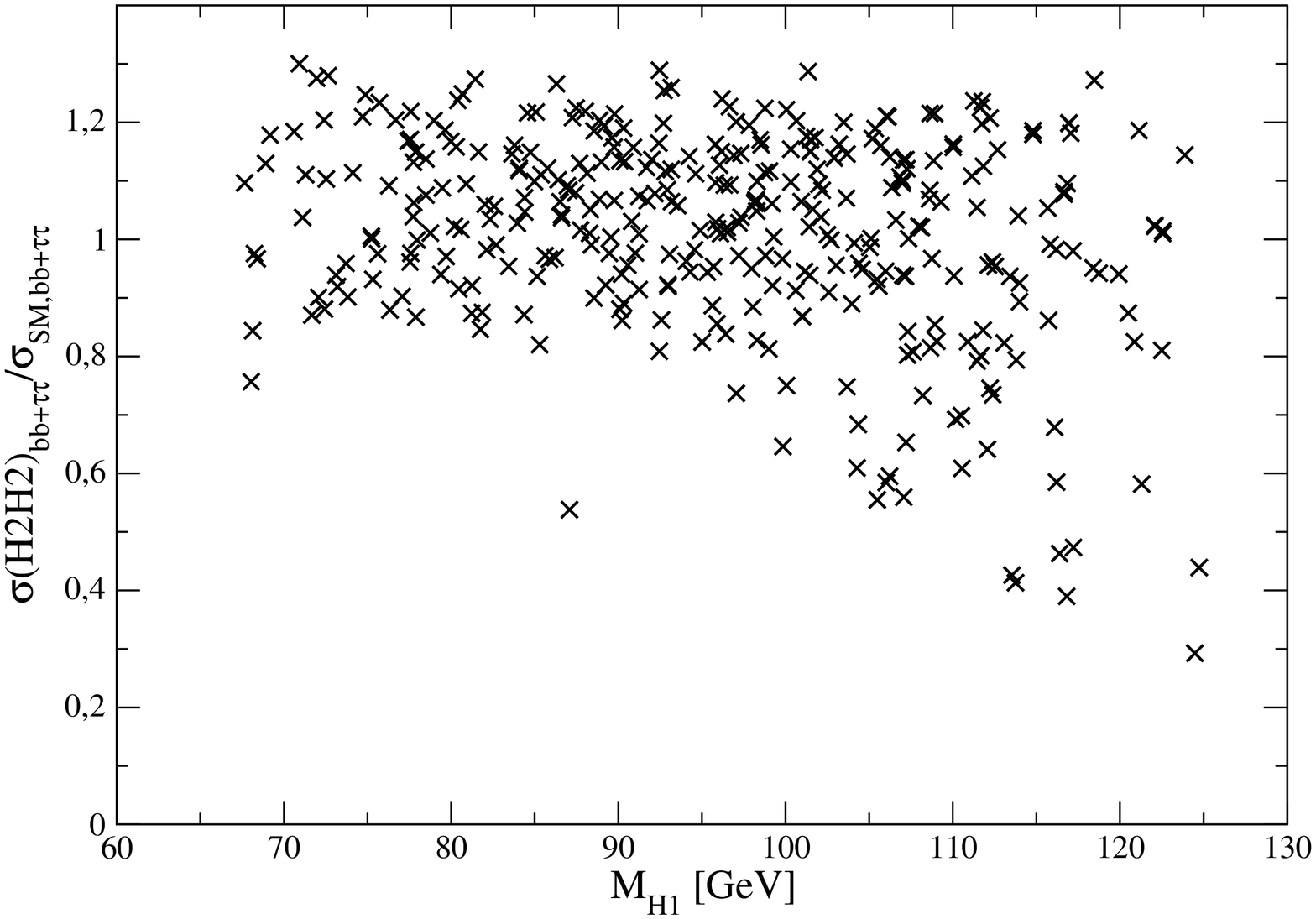}\quad
\includegraphics[width=6.1cm,angle=-90]{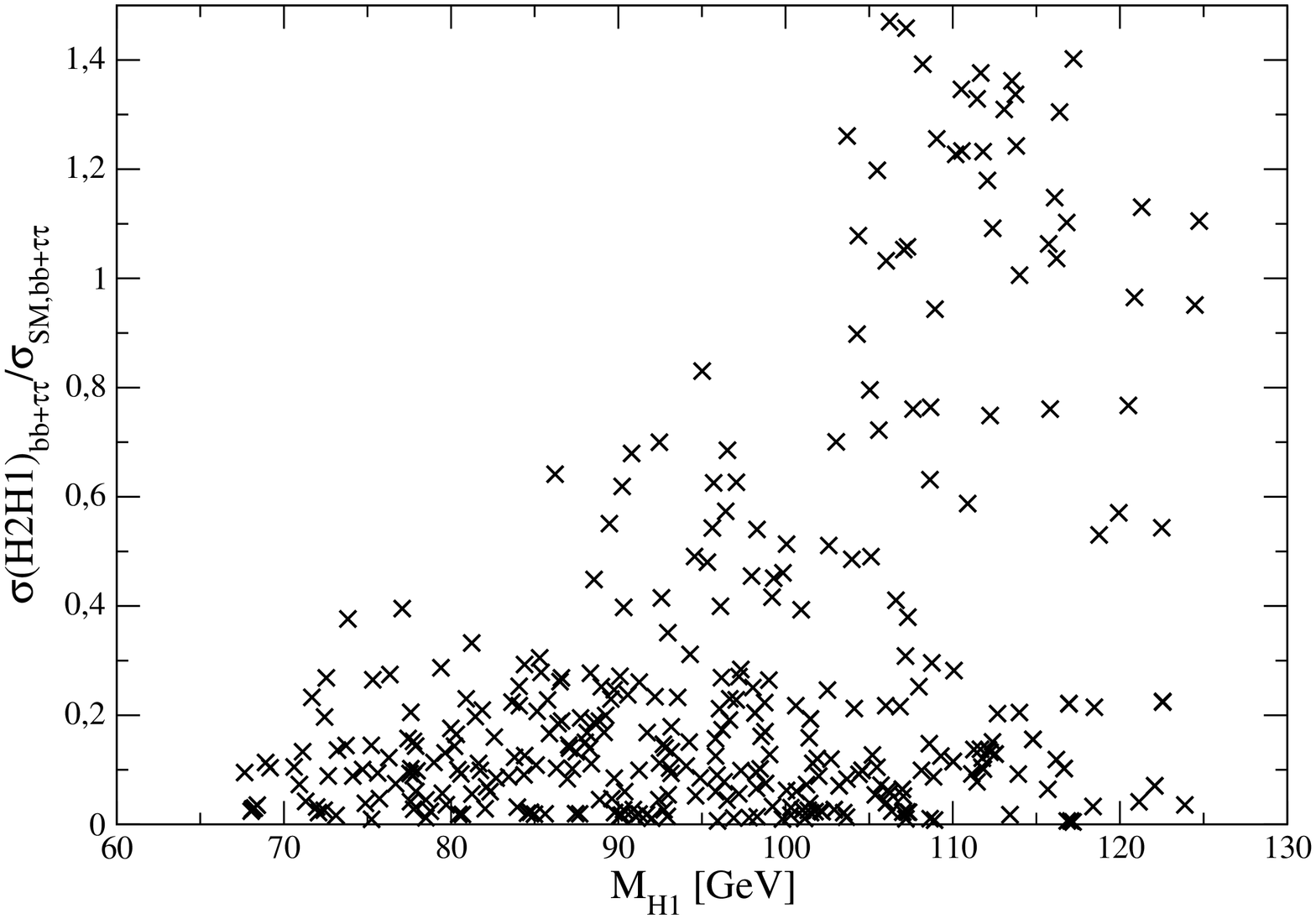}\\
\includegraphics[width=6.1cm,angle=-90]{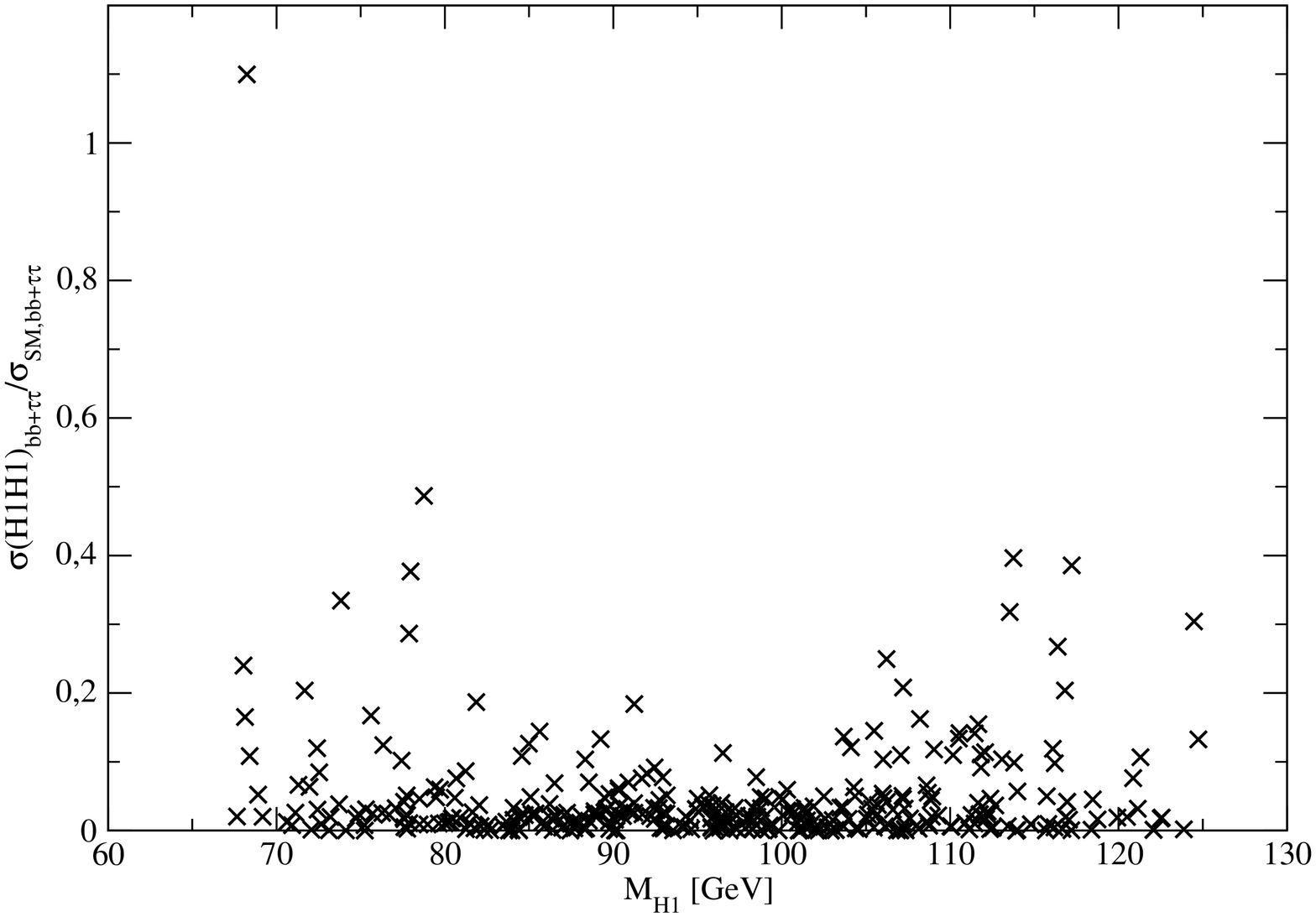}\quad
\includegraphics[width=6.1cm,angle=-90]{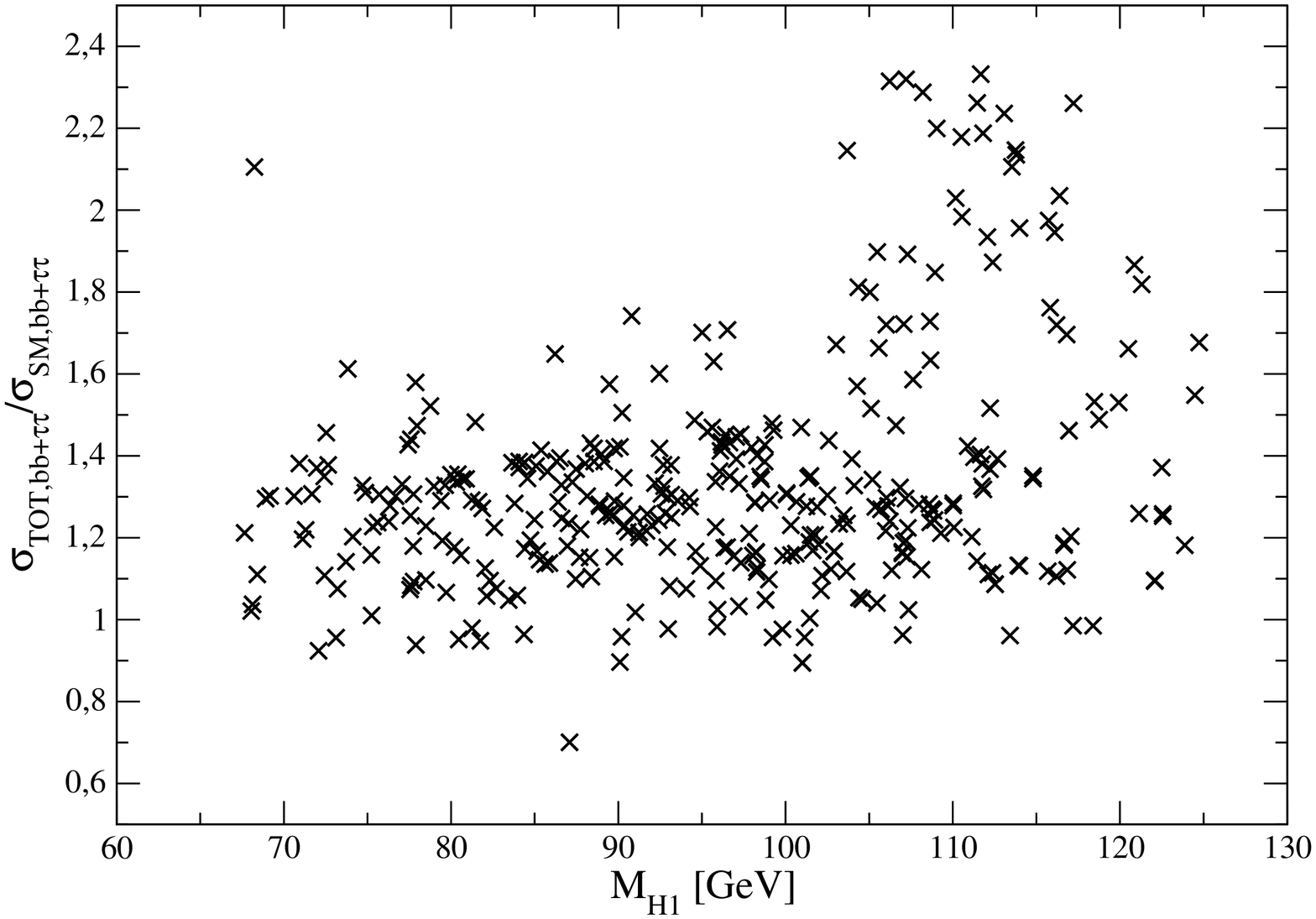}
\caption{The Higgs pair production cross sections relative to the SM for
the final states $H_2+H_2 \to b\bar{b}+\tau^+\tau^-$ (top left),
$H_2+H_1 \to b\bar{b}+\tau^+\tau^-$ (top right), $H_1+H_1 \to
b\bar{b}+\tau^+\tau^-$ (bottom left) and the sum over all $H_i$ into
$b\bar{b}+\tau^+\tau^-$ (bottom right).}
\label{fig:3}
\end{figure}

Although the cross sections weighted by the branching fractions relative
to the SM differ in general from the full relative cross sections shown
in Fig.~2 for
given points in parameter space, the general pattern is similar: the sum
over cross sections for all Higgs pairs is practically never below the
SM value, and can be amplified by a factor $\sim 2.4$. This originates
from possible $H_2+H_1 \to b\bar{b}+\tau^+\tau^-$ processes, whose cross
sections can reach beyond-the-SM values by itself.

Unfortunately the production of the lighter Higgs boson $H_1$ in either
the $H_2+H_1 \to b\bar{b}+\tau^+\tau^-$ or $H_1+H_1 \to
b\bar{b}+\tau^+\tau^-$ final state does not always have a cross section
of the order the SM Higgs pair production. Hence the detection of $H_1$
in this channel is a spectacular possibility, but not guaranteed.

\begin{figure}[ht!]\centering
\includegraphics[width=6.1cm,angle=-90]{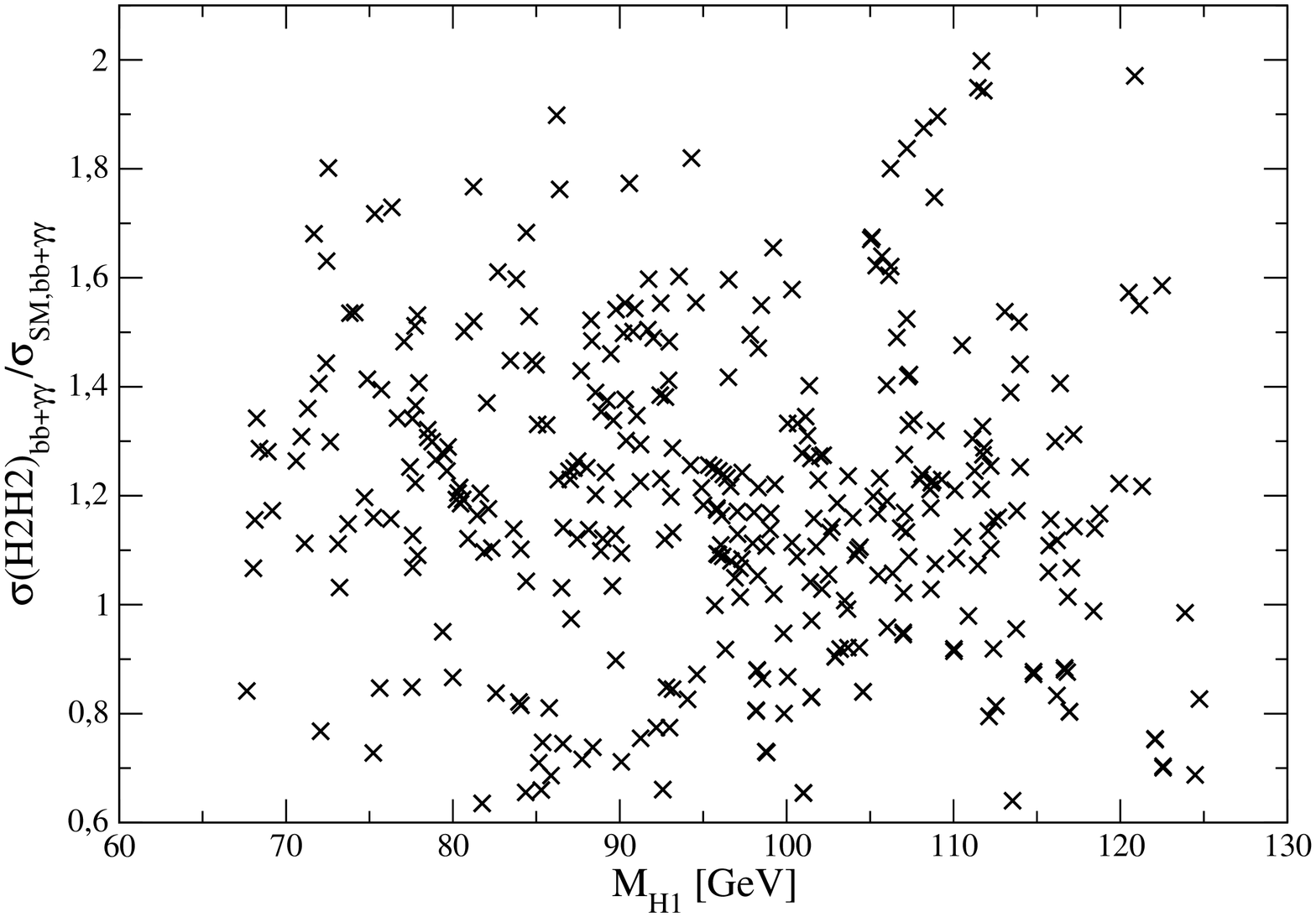}\quad
\includegraphics[width=6.1cm,angle=-90]{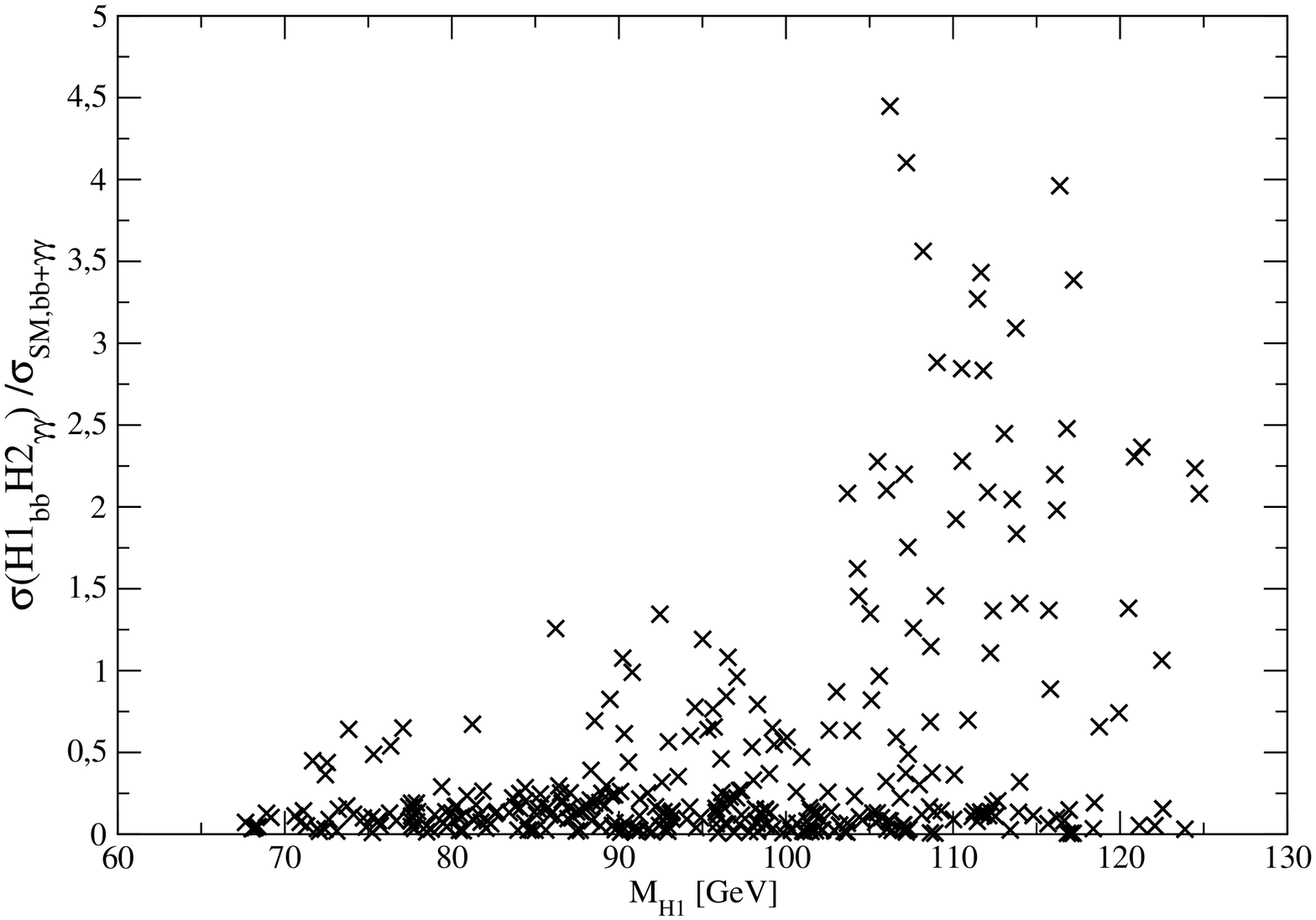}\\
\includegraphics[width=6.1cm,angle=-90]{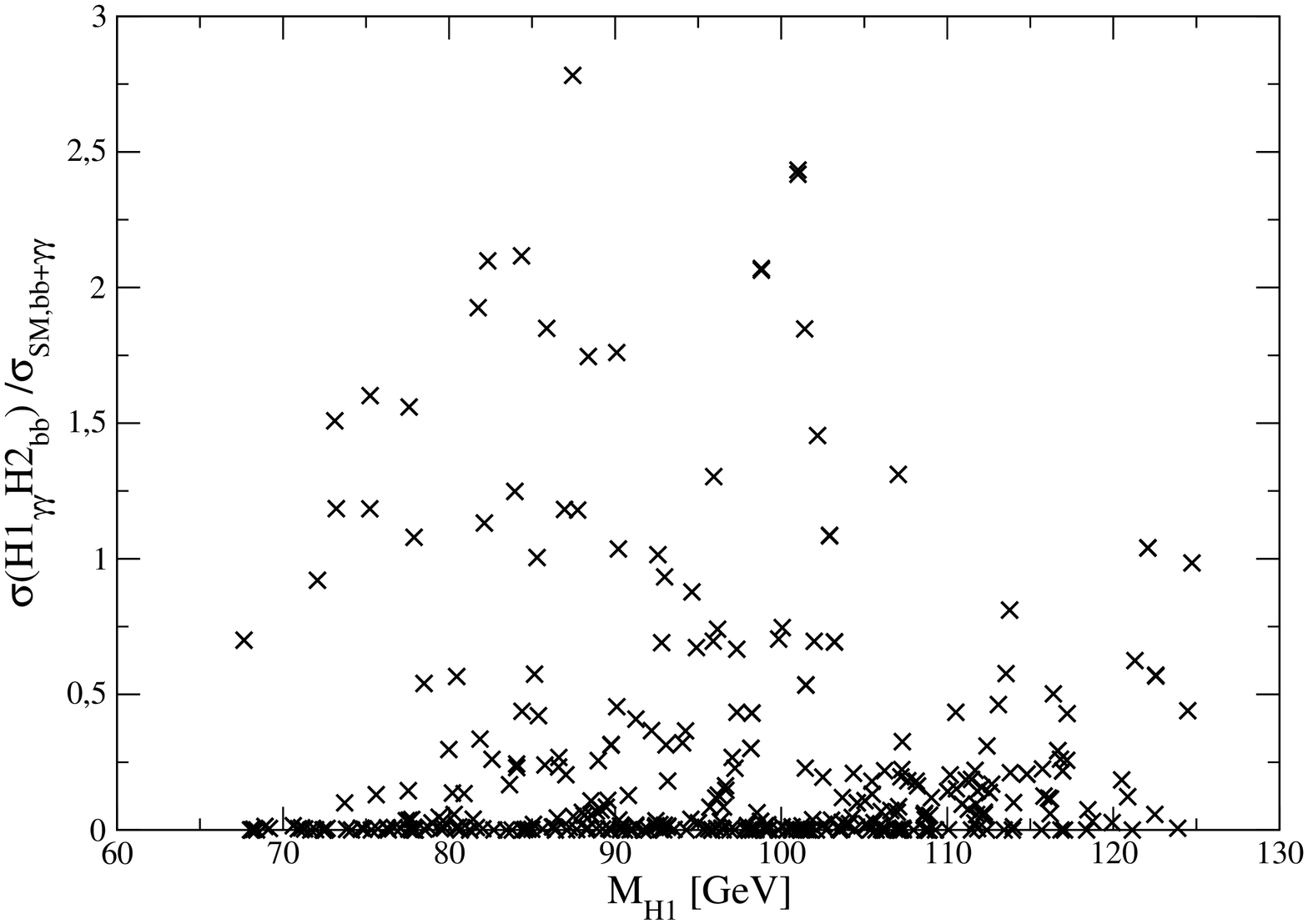}\quad
\includegraphics[width=6.1cm,angle=-90]{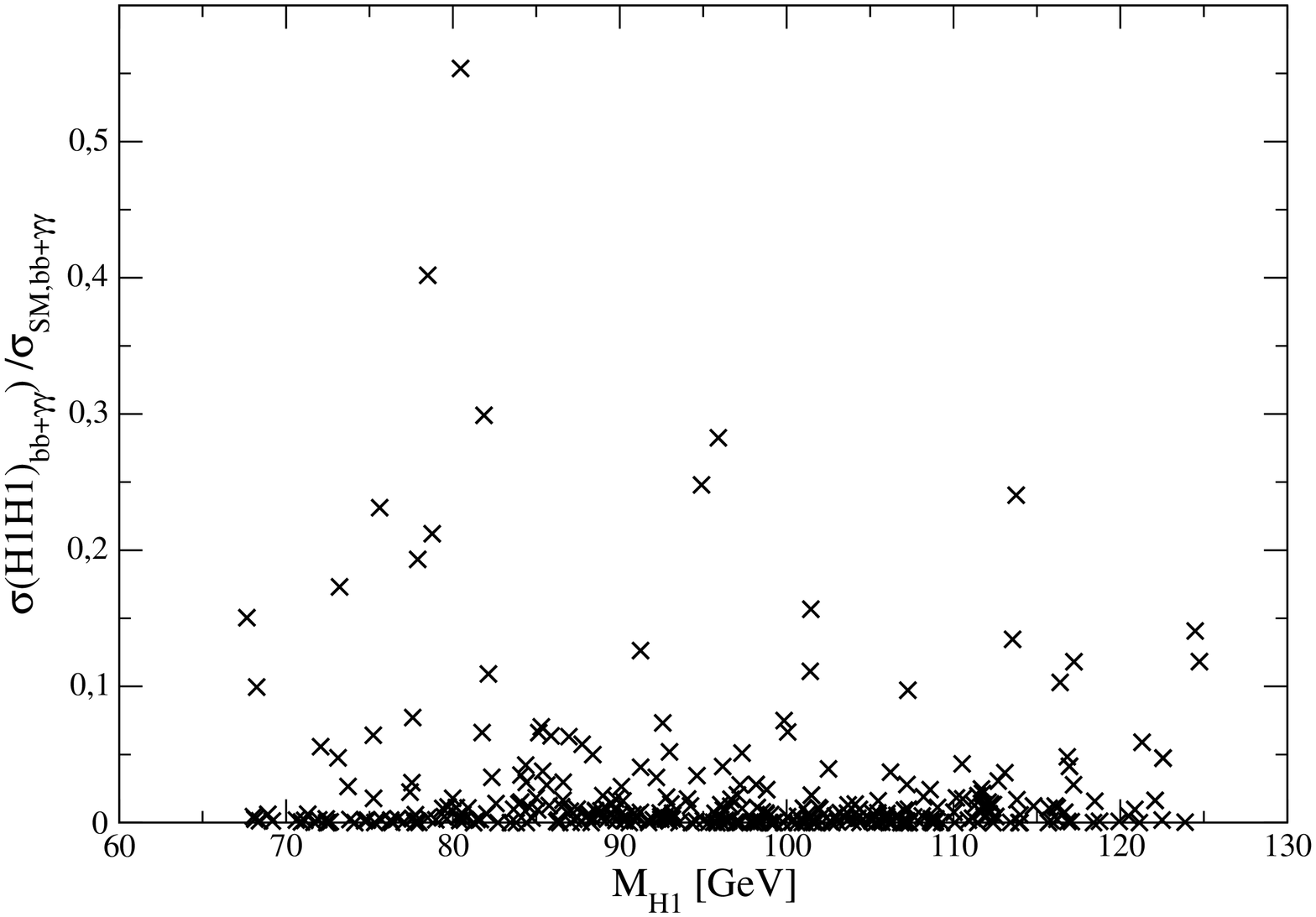}\\
\includegraphics[width=6.1cm,angle=-90]{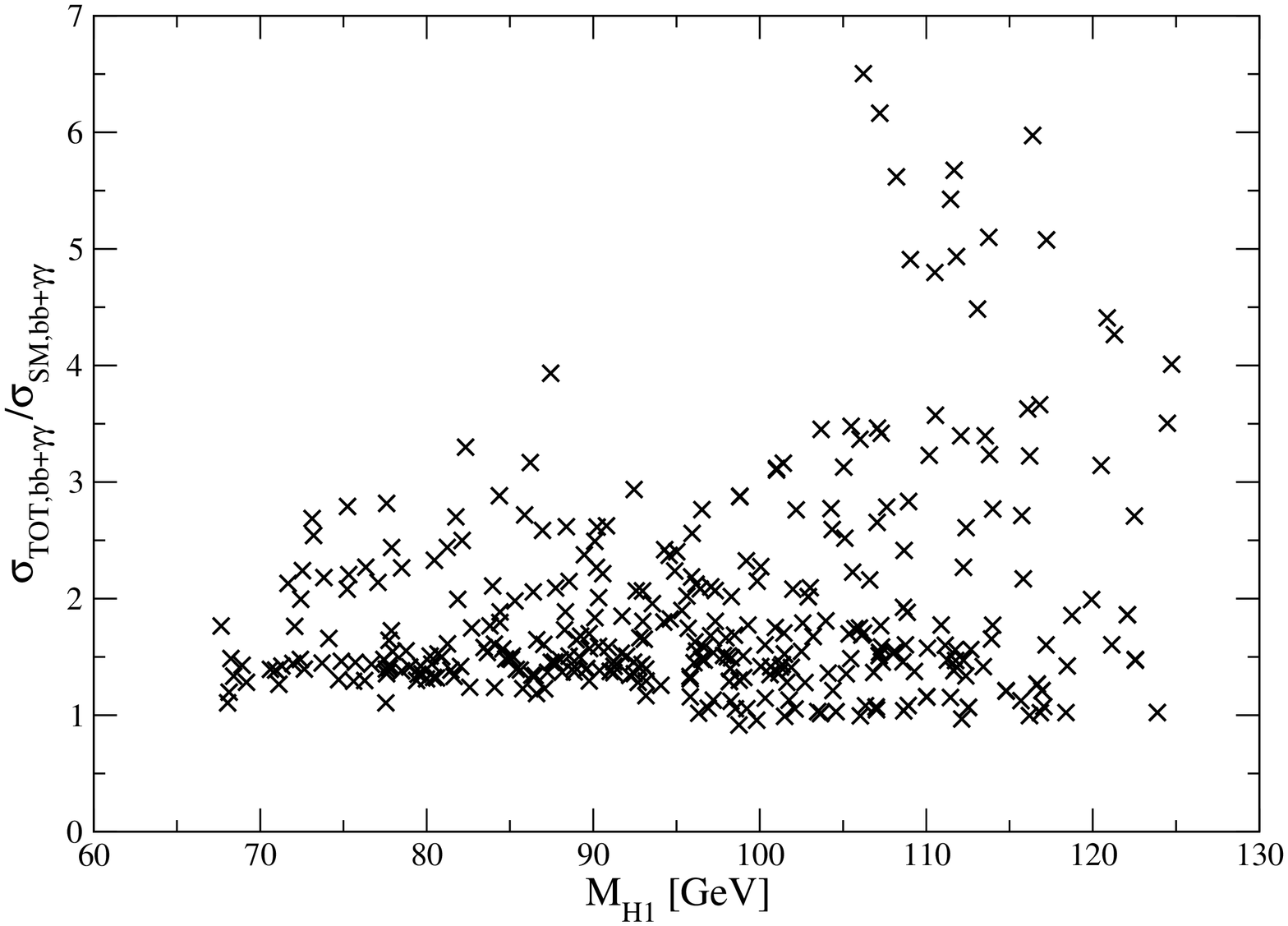}
\caption{The Higgs pair production cross sections relative to the SM for
the final states $H_2+H_2 \to b\bar{b}+\gamma\gamma$ (top left), $(H_1
\to b\bar{b})+(H_2\to \gamma\gamma)$ (top right), $(H_1 \to
\gamma\gamma) +(H_2\to b\bar{b})$ (middle left), $H_1+H_1 \to
b\bar{b}+\gamma\gamma$ (middle right) and the sum over all $H_i$ into
$b\bar{b}+\gamma\gamma$ (bottom).}
\label{fig:4}
\end{figure}

Another potential Higgs pair discovery channel is the
$b\bar{b}+\gamma\gamma$ final state \cite{Baur:2003gp}. Again we
consider the production cross sections multiplied by the branching
fractions relative to the ones of a 125~GeV SM-like Higgs boson, which
would be the target once such searches will be performed. In this case
the final states $(H_1 \to b\bar{b})+ (H_2 \to \gamma\gamma)$ and
$(H_2 \to b\bar{b})+ (H_1 \to \gamma\gamma)$ have to be considered
separately. The relative cross sections times branching fractions for
all possible distinct final states together with their sum
are shown in Fig.~4.

In Fig.~4 we see cases of spectacular enhancements with respect to the
SM cross section in this channel. In the final state $(H_1\to b\bar{b})
+ (H_2 \to \gamma\gamma)$ (up to a factor $\sim 4.5$, top right) these
originate from an enhancement of the branching fraction for $H_2 \to
\gamma\gamma$ by a factor up to $\sim 2.3$, consistent with the present
measurements of signal rates at the LHC due to a slightly reduced $H_2$
production cross section, together with enhancements of the branching
fraction for $H_1 \to b\bar{b}$ and the production cross section into
$H_2+H_1$. Due to the absence of the latter factors and a reduction of
the branching fraction for $H_2 \to b\bar{b}$, the enhancements of the
cross sections in the final state $H_2+H_2 \to b\bar{b}+\gamma\gamma$
for the same points in the top left of Fig.~4 are less spectacular, but
lead to relative signal rates up to a factor $\sim 6.5$ larger for the
sum over the final states $H_1+H_2$ and $H_2+H_2$. (These final states
differ, however, in the invariant mass of the $b\bar{b}$ system
corresponding to $M_{H_1}$ or $M_{H_2}$, respectively.)

Also in the distinguishable final state $(H_1\to \gamma\gamma) +(H_2 \to
b\bar{b})$ enhancements up to a factor $\sim 2.8$ are possible. These
originate from enhancements of the branching fraction for $H_1 \to
\gamma\gamma$, a possibility pointed out in \cite{Ellwanger:2010nf}. On
the other hand, as before it is not guaranteed that any cross section
involving $H_1$ is measurably large; one just ends up with a ``no-lose
theorem'' stating that at least the cross section for the sum over all
$H_1$, $H_2$ final states never falls below the SM cross section.

\section{Conclusions}

The regions in the NMSSM parameter space giving rise to another mostly
singlet-like Higgs boson $H_1$ with a mass below 125~GeV are
particularly natural, since here mixing effects lead to an increase of
the mass of the mostly SM-like state $H_2$. In the present work we
studied the possible impact of such mixings on the Higgs pair production
cross sections in gluon fusion at 14~TeV and the $b\bar{b}+\tau^+\tau^-$
and $b\bar{b}+\gamma\gamma$ final states. A priori it was not clear
whether, even after summing over the $H_1$, $H_2$ final states, the
cross sections are always as large as in the SM for a single 125~GeV
Higgs boson. Our results indicate that this is the case. Cross sections
larger than in the SM are possible, notably in the
$b\bar{b}+\gamma\gamma$ final state involving both $H_1$ and $H_2$. But
unfortunately it is not guaranteed that the additional state $H_1$ has
always a sufficiently large production rate in Higgs pair production to
be detectable; in this case one would have to rely on either its direct
production, or on decays of the heavier state $H_3$ into $H_1$.

On the other hand, given the possibility that $H_1$ is visible only in
Higgs pair production processes, sufficiently flexible cuts should be
applied in future analyses in order not to miss it even for
$M_{H_1}$ well below 125~GeV.

\section*{Acknowledgements}

We acknowledge support from the French
ANR~LFV-CPV-LHC, ANR~STR-COSMO, the European Union FP7 ITN INVISIBLES
(Marie Curie Actions,~PITN-GA-2011-289442) and the ERC advanced grant
Higgs@LHC.

\end{document}